\documentclass[amssymb,amsmath,floatfix,aps,pre,twocolumn,superscriptaddress,longbibliography, nofootinbib]{revtex4-2}
\usepackage[latin1]{inputenc}

\usepackage{mathptmx}
\usepackage[T1]{fontenc}

\usepackage{graphicx}
\usepackage{amsmath}
\usepackage{amssymb}
\usepackage{amsfonts}
\usepackage[scr=rsfs]{mathalpha}
\usepackage{bm}
\usepackage{xr-hyper}
\usepackage{hyperref}
\IfFileExists{SM.aux}{\externaldocument[SM-]{SM}}{\IfFileExists{SM-xr.tex}{\AtBeginDocument{\newlabel{SM-FirstPage}{{}{1}{}{section*.1}{}}
\newlabel{SM-sec:potential}{{I}{1}{}{section*.3}{}}
\newlabel{SM-eq:potential}{{S2}{1}{}{equation.1.2}{}}
\newlabel{SM-eq:potential2}{{S5}{1}{}{equation.1.5}{}}
\newlabel{SM-sec:optimalprot}{{II}{1}{}{section*.4}{}}
\newlabel{SM-eq:Alamsol}{{S10}{2}{}{equation.2.10}{}}
\newlabel{SM-eq:newcost}{{S11}{2}{}{equation.2.11}{}}
\newlabel{SM-eq:Hamquad}{{S12}{2}{}{equation.2.12}{}}
\newlabel{SM-eq:hameom}{{S13}{2}{}{equation.2.13}{}}
\newlabel{SM-eq:mueqs}{{S15}{2}{}{equation.2.15}{}}
\newlabel{SM-eq:relation}{{S18}{2}{}{equation.2.18}{}}
\newlabel{SM-eq:Wtf}{{S20}{2}{}{equation.2.20}{}}
\newlabel{SM-sec:tcrit}{{III}{3}{}{section*.5}{}}
\newlabel{SM-eq:Wexp}{{S21}{3}{}{equation.3.21}{}}
\newlabel{SM-sec:ftdpts}{{IV}{3}{}{section*.6}{}}
\newlabel{SM-eq:action}{{S23}{3}{}{equation.4.23}{}}
\newlabel{SM-eq:Vtf}{{S27}{3}{}{equation.4.27}{}}
\newlabel{SM-eq:AV0t}{{S28}{3}{}{equation.4.28}{}}
\newlabel{SM-sec:Aexpmeth}{{V}{4}{}{section*.7}{}}
\newlabel{SM-sec:Aexpsetup}{{V\,A}{4}{}{section*.8}{}}
\newlabel{SM-sec:Aexpmeas}{{V\,B}{4}{}{section*.9}{}}
\newlabel{SM-sec:Aexpcontrol}{{V\,B\,1}{4}{}{section*.10}{}}
\newlabel{SM-eq:Wcrooks}{{S31}{4}{}{equation.5.31}{}}
\newlabel{SM-sec:Aexpfreerelax}{{V\,B\,2}{5}{}{section*.11}{}}
\newlabel{SM-eq:probintegral}{{S32}{5}{}{equation.5.32}{}}
\newlabel{SM-eq:boltzmann}{{S33}{5}{}{equation.5.33}{}}
\newlabel{SM-eq:minrel}{{S36}{5}{}{equation.5.36}{}}
\newlabel{SM-eq:minrel2}{{S37}{5}{}{equation.5.37}{}}
\newlabel{SM-LastBibItem}{{3}{6}{}{section*.12}{}}
\newlabel{SM-LastPage}{{}{6}{}{}{}}
}}{}}
\usepackage{bbold}
\usepackage{xcolor}
\usepackage{overpic}

\usepackage{siunitx}
\newcommand{\algn}[1]{\begin{align} #1 \end{align}}

\newcommand{\ed}{\ensuremath{\text{d}}}

\newcommand{\ms}[1]{\ensuremath{\mathscr{#1}}}

\newcommand{\kb}{\ensuremath{k_\text{B}}}
\newcommand{\eqnlab}[1]{\label{eq:#1}}

\newcommand{\figlab}[1]{\label{fig:#1}}
\newcommand{\eqnref}[1]{\eqref{eq:#1}}
\newcommand{\Eqnref}[1]{Eq.~\eqref{eq:#1}}
\newcommand{\Eqsref}[1]{Eqs.~\eqref{eq:#1}}
\newcommand{\secref}[1]{\ref{SM-sec:#1}}
\newcommand{\Secref}[1]{Sec.~\ref{SM-sec:#1}}
\newcommand{\figref}[1]{\ref{fig:#1}}
\newcommand{\Figref}[1]{Fig.~\ref{fig:#1}}
\newcommand{\Figsref}[1]{Figs.~\ref{fig:#1}}

\begin{document}
\title{Finite-time transitions in optimal control and non-equilibrium relaxation}

\author{Jan Meibohm}
\affiliation{Institute for Physics and Astronomy, Technische Universit\"at Berlin, Hardenbergstra\ss{}e 36, 10623 Berlin}
\author{Samuel Monter}
\affiliation{Faculty of Physics, University of Konstanz, Konstanz, Germany}
\author{Sarah A. M. Loos}
\affiliation{Max Planck Institute for Dynamics and Self-Organization, G\"ottingen}
\author{Clemens Bechinger}
\affiliation{Faculty of Physics, University of Konstanz, Konstanz, Germany}

\begin{abstract}
We theoretically and experimentally study finite-time optimal control of a colloidal particle steered through a spatially inhomogeneous environment, modeled by a position-dependent energetic cost at the final state. The competition between this state-dependent penalty and path-dependent dissipation gives rise to a sharp transition in the control strategy at a critical control duration. We further show that this transition is linked to a dynamical phase transition in nonequilibrium relaxation after a quench, where the control cost maps onto the rate function governing rare trajectories.
\end{abstract}

\maketitle

\paragraph*{Introduction.}
With ongoing miniaturization of technological devices, the optimization of noisy processes has become increasingly important. At small scales, strong fluctuations are unavoidable, and performance is constrained by energy dissipation. Similar constraints govern biological systems, such as molecular motors and microorganisms, which must operate efficiently under strong noise and finite-time conditions~\cite{Chi24,Ju25,How01,Jul97,Sog17}. A defining feature of such processes is that they occur over finite durations, leading to a fundamental trade-off between energetic cost and speed: faster protocols incur higher dissipation, whereas slower protocols reduce energetic cost at the expense of performance. Optimal control theory provides a natural framework to quantify and optimize such trade-off by identifying protocols that minimize a prescribed cost functional over a finite-time horizon~\cite{Fle75,Bec21,Alv25}.

Even in simple settings, such as dragging colloidal particles or navigating active swimmers through fluctuating environments, optimal protocols can be nontrivial, with discontinuities arising at the protocol boundaries~\cite{Sch07,Loo24,Col17,Gar25,Gup23}. In structured environments with spatially varying energetic cost, optimal control becomes further enriched: the optimization must specify not only how fast to drive the system, but also where to go, since the location of the endpoint is part of the optimization task. This competition between path-dependent dissipation and position-dependent energetic penalty gives rise to distinct optimal strategies.

Here we show that this trade-off leads to a finite-time transition between qualitatively distinct optimal protocols. For a broad class of control problems with nonconvex energetic penalties, we demonstrate theoretically and experimentally that a critical protocol duration separates a regime where it is optimal to accept an energetic penalty from one in which active steering towards less costly regions is favorable. For symmetric landscapes, this transition is accompanied by spontaneous symmetry breaking~\cite{Kap05a,Kap05b} of the optimal solution, reminiscent of a continuous phase transition. 

Exploiting the connection between stochastic optimal control and large deviation theory~\cite{Fle06,Tou09,Che15b,Gra19}, we further show that this control transition can be mapped onto a finite-time dynamical phase transition~\cite{Mei22a,Blo22,Mei23b,Vad24,Asr26} (FTDPT) in the relaxation dynamics of a freely diffusing stochastic system. In this mapping, the optimal control cost plays the role of a rate function, and the transition manifests as a nonanalyticity associated with a change in the dominant relaxation pathway, from rare fluctuations at short times to typical relaxation trajectories at longer times. While direct observation of such rare events requires exponentially many realizations, optimal control experiments access the same information through averages over controlled trajectories, thereby bypassing this sampling bottleneck. Our results thus establish optimal control as an experimentally accessible route to dynamical phase transitions and, more broadly, to rare event physics in mesoscopic systems. In a companion paper~\cite{Mei26c}, we show that the finite-time control transition and its direct link to FTDPTs generalize to a much broader class of control problems.

\paragraph*{Control problem.}
\begin{figure}
	\includegraphics[width =.9\linewidth]{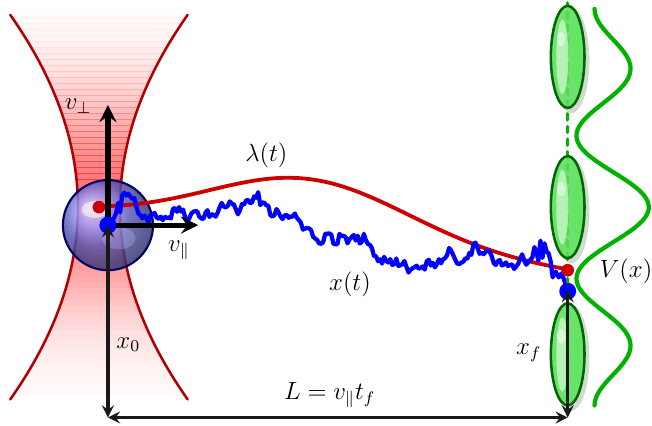}
	\caption{A particle (blue) is driven by a harmonic trap (red) at constant mean horizontal velocity $v_\parallel$ toward an obstacle (green), characterized by an energetic cost $V(x)$. The trap controls the lateral velocity $v_\perp$ to minimize the total work performed on the particle.}\figlab{sketch}
\end{figure}

As a paradigmatic control scenario in a heterogeneous environment, we here consider the situation sketched in \Figref{sketch}. A microscopic particle (blue), suspended in a viscous fluid and subjected to thermal noise, is steered by a harmonic trap (red) whose time-dependent position $\lambda(t)$ serves as control parameter. Starting in equilibrium at time $t=0$ at position $x_0$, the particle moves with constant mean longitudinal velocity $v_\parallel$ to the right and reaches an obstacle (green) after time $t_f$ at distance $L=v_\parallel t_f$.
The obstacle is modeled by a structured energy landscape $V(x)$ extending in the transverse direction with spatially varying cost. The average particle motion is fully determined by the trap position. While its longitudinal velocity $v_\parallel$ is fixed, the transverse particle velocity $v_\perp(t)$ is controlled to minimize a total cost comprising an energetic penalty $V(x_f)$ and the dissipated energy accumulated along the path.

As an illustrative example, we consider an overdamped particle dynamics in presence of
additive white noise $\xi(t)$, as described by the following Langevin equation
\algn{\eqnlab{Langevinequ}
    \dot x(t) = -\tau_\text{p}^{-1}[x(t)-\lambda(t)] + \sqrt{2\kb T/\gamma}\,\xi(t)\,,
}
which provides an accurate minimal model of our experimental setup.
Here, $T$ denotes the ambient fluid's temperature, $\kb$ the Boltzmann constant, and $\tau_\text{p}=\gamma/\kappa$ the particle relaxation time, determined by the particle's friction coefficient $\gamma$ and the optical trap stiffness $\kappa$. The work required for steering the particle in perpendicular direction is given by $W^s_{t_f} = \kappa\int_0^{t_f}\!\!\ed t\ \dot\lambda(\lambda-x)$~\cite{Sei12,Pel21}. The objective is to minimize the mean $\langle W_{t_f}\rangle$ of the total work $W_{t_f} = W^s_{t_f}+V(x_f)$ over both the control protocol
$\lambda(t)$ and the mean particle position $u(t) = \langle x(t)\rangle$, for a fixed mean initial
condition $u_0=\langle x(0)\rangle$. The mean $\langle W_{t_f}\rangle$ reads
\algn{\eqnlab{cost}
    \ms{W}_{t_f}[u(t),\lambda(t)] = \langle W_{t_f} \rangle =
    \kappa\int_0^{t_f}\!\!\ed t \, \dot\lambda(t)[\lambda(t)-u(t)] + \tilde V(u_f)\,,
}
where $u_f = \langle x(t_f)\rangle$ is the final mean position and $\tilde V(u_f)=\langle V(x_f) \rangle$ is the noise-averaged penalty. We constrain the minimization of $\ms{W}_{t_f}$ so that $u(t)$ satisfies the noise-averaged dynamics \Eqnref{Langevinequ}, by introducing a Lagrange multiplier.
Applying the Pontryagin principle~\cite{Fle75,Bec21} then yields a quadratic control Hamiltonian (Sec.~\secref{optimalprot} in~\cite{sup}), whose optimal solution gives
\algn{
	\ms{W}_{t_f}(u_f,u_0) &= \frac{\gamma(u_f-u_0)^2}{t_f} + \tilde V(u_f)\,, \eqnlab{cc3}
}
with $\lambda_0=u_0$ and $\lambda_f=u_f$, reflecting that the system does not store excess potential energy at $t_f$. These boundary terms are not imposed, but are a direct result of the minimization.
The first term on the r.h.s. of \Eqnref{cc3} accounts for the total dissipative costs and corresponds to the obstacle-free solution~\cite{Sch07}, while the second term accounts for the mean penalty at $u_f$. 

In a final step, we optimize $\ms{W}_{t_f}$ over $u_f$, i.e.,
\algn{\eqnlab{Cmin}
	\ms{W}^*_{t_f}(u_0) = \min_{u_f} \ms{W}_{t_f}(u_f,u_0)\,,
}
to obtain the optimal total cost $\ms{W}^*_{t_f}(u_0)= \ms{W}_{t_f}(u^*_f,u_0)$, which solely depends on the initial particle position $u_0$. Here we denote by $u^*_f(u_0,t_f) = \text{argmin}_{u_f}\ms{W}_{t_f}(u_f,u_0)$ the associated minimizer.
\paragraph*{Finite-time control transition.}
Examination of the cost function in such structured environment leads to a general observation: The competition between the position-dependent penalty $V(x_f)$ and viscous dissipation penalizing large displacements along $x$. As we will show below, this competition can lead to abrupt switches in the optimal control strategies upon changing of the overall process duration. To illustrate this, we consider a symmetric penalty function $ V(x_f)=V(-x_f)$ with $V'(0) = 0$ and $V''(0) < 0$. Setting the initial particle position to $x_0 = \lambda_0 = 0$, a \emph{zero-control} solution [$v_\perp(t)=0$] yields the maximal penalty since the particle moves towards the maximum of $V(x)$. From \Eqsref{cc3}, it follows that $u_f = 0$ satisfies $\partial_{u_f}\ms{W}_{t_f} = 0$, and is therefore a stationary point of $\ms{W}_{t_f}$.

To assess its stability, we evaluate the second derivative
\algn{\eqnlab{ddW}
\partial^2_{u_f}\ms{W}_{t_f}(0,0) = \frac{2\gamma}{t_f} + \tilde V''(0).
}
For short protocol durations $t_f \ll \tau_{\text{p}}$, the dissipative contribution dominates, rendering any displacement too costly and stabilizing the zero-control solution: $\lambda^\ast(t)=u^\ast(t)=0$. However, when $t_f$ exceeds a critical time
\algn{\eqnlab{tcrit}
t_\text{c} = \frac{2\gamma}{-\tilde V''(0)},
}
dissipation becomes sufficiently weak that the energetic penalty dominates, see \Secref{tcrit} in \cite{sup}. Consequently, \Eqnref{ddW} changes sign and the zero-control solution becomes suboptimal.

As a consequence, for $t_f > t_\text{c}$ a new strategy emerges. Now, it is favorable to steer the particle towards regions of lower penalty, so the optimal protocol is in general given by
\algn{\eqnlab{lamsol}
\lambda^*(t) =\begin{cases}
u_0, & t=0,\\
t_f^{-1}(u_{f}^* - u_0)\left(t+\tau_\text{p}\right) + u_0\,,& 0 < t < t_f\,,\\
u^\ast_f& t=t_f\,.
\end{cases}
}
Thus, $\lambda^\ast(t)$ becomes linear in time with discontinuities at the boundaries, similar to the optimal control without penalty~\cite{Sch07}. Recall that the final position $\lambda_f^* = u_f^*$ is determined by minimizing the mean total energetic cost through \Eqnref{Cmin}.
In the limit $u_0 \to 0$ and $t_f \to t_\text{c}$, we recover $\lambda^*(t) \to 0$, such that the linear solution connects continuously to the zero-control protocol.

It is useful to define the magnitude of the optimal final particle position $|u_f^*|$ as an order parameter for the transition, as it directly captures the onset of symmetry breaking. For a symmetric and locally concave penalty function, where $\tilde V(u_f)$ is even around $u_f = 0$ and $\tilde V''(0) < 0$, the competition between dissipation and energetic cost leads to a bifurcation of the optimal solution. For $t_f < t_\text{c}$, the symmetric solution $u_f^* = 0$ is stable, whereas for $t_f > t_\text{c}$, two symmetry-broken solutions with $u_f^* \neq 0$ emerge.

This behavior admits a direct interpretation in terms of a continuous phase transition. The control time $t_f$ acts as a tuning parameter that shifts the balance between dissipation and energetic cost, leading to a loss of stability of the symmetric solution. At the critical time $t_\text{c}$, the curvature of $\ms{W}_{t_f}$ at $u_f = 0$ changes sign, analogous to the vanishing quadratic coefficient in a Landau expansion~\cite{Cha95}.

\paragraph*{Experimental results.}
We now turn to an experimental test of these predictions, employing a colloidal silica particle (diameter $\approx 2.73\,\unit{\micro\meter}$) that is dragged through a viscous fluid by optical tweezers. The fluid is confined in a sample cell with height $100\,\unit{\mu m}$ and consists of a 1:1 (volume fraction) mixture of water and glycerol, resulting in friction coefficients $\gamma = 0.11\,\unit{\mu Ns/m}$ in the control experiment and $\gamma = 0.18\,\unit{\mu Ns/m}$ in the relaxation experiment. The tweezers are created by a focused $532\,\unit{\nano\meter}$ laser beam that generates a harmonic trapping potential
\begin{equation}\eqnlab{Upot}
	U(x,\lambda) = \frac{\kappa}{2}(x-\lambda)^2\,.
\end{equation}
For trap stiffnesses in the range $\kappa = (0.51$--$0.62) \pm 0.05\,\unit{\micro\newton\per\meter}$, as measured across different experimental runs and consistent with the 5\,\% polydispersity of the particles, the relaxation time in \Eqnref{Langevinequ} was found to range between $\tau_\text{p} = (0.172$--$0.216) \pm 0.02\,\unit{\second}$. Temporal control of $\lambda$ was realized by deflecting the laser beam with an acousto-optical deflector (AOD), enabling dynamic positioning of the optical trap center with a spatial accuracy of 10~\unit{\nano\meter} and a repetition rate of $10^{-4}~\unit{\second}$. Particle positions are tracked using digital video microscopy with a spatial and temporal resolution of $10\,\unit{\nano\meter}$ and $2.5-4\,\unit{\milli\second}$, respectively. The sample was kept at a constant temperature of 25~\unit{\celsius}. For details see \Secref{Aexpmeth} in~\cite{sup}.

In our experiments, we consider a penalty landscape following a symmetric double-well potential, which can be interpreted as a soft obstacle with a soft double slit,
\begin{equation}\eqnlab{Vx}
V(x) = \frac{V_0}{4} \left(\frac{x^2}{x_\text{m}^2} - 1\right)^2\,,
\end{equation}
with openings at $\pm x_\text{m}$, where $x_\text{m} = 1\,\unit{\micro\meter}$ and $V_0 = \kappa x_\text{m}^2$ in the control experiment. The relaxation experiment below uses the same form with $x_\text{m} = 0.2\,\unit{\micro\meter}$.

For this choice of $V(x)$, we calculated the optimal protocols $\lambda^*(t)$ and implemented them in our experiments (see Secs.~\secref{optimalprot} and \secref{Aexpmeth} in \cite{sup}). The dashed lines in \Figsref{protocols}(a) and \figref{protocols}(b) show $\lambda^*(t)$ for $t_f < t_\text{c} = 0.36 \pm 0.04\,\unit{\second}$ and $t_f > t_\text{c}$, respectively, for different initial trap positions $\lambda_0 = u_0$ along the $x$-direction.  

For $t_f < t_\text{c}$ [Fig.~2(a)], the zero-control solution [$\lambda^*(t) = u^*(t) = 0$] is optimal, if $u_0 = 0$, accepting a final penalty $V(0) = V_0/4$ (green). For non-zero $u_0$, the optimal protocol continuously evolves into solutions with finite perpendicular motion, which exhibit discontinuous jumps at the beginning and end of the protocol, similar to optimal transport without a penalty~\cite{Sch07}. By contrast, for $t_f > t_\text{c}$ [Fig.~2(b)], such discontinuities occur already at $u_0 = 0$, and the protocol transports the particle towards either $\pm x_\text{m}$. 
Only for the special case that $u_0$ coincides with a minimum of $\tilde V$,
the zero-control solution remains optimal for all protocol durations $t_f$.

Based on the measured particle trajectories [red and blue lines in Fig.~2(a,b)] and the penalty $V(x_f)$, we compute $\ms{W}^*_{t_f}$ from the discretized integral in \Eqnref{cost} for 80 experimental realizations of each parameter set $(u_0, t_f)$. Figure~\figref{protocols}(c) shows the measured $\ms{W}^*_{t_f}=\langle W_{t_f}\rangle$ (symbols) as a function of $t_f$ for $u_0 = 0$ together with the theory [\Eqnref{Cmin}, solid line]. For $t_f < t_\text{c}$, the zero-control solution is optimal, so $\ms{W}^*_{t_f}=\tilde V(0)$ remains constant. For $t_f > t_\text{c}$, by contrast, the non-trivial optimal protocol~\eqnref{lamsol} reduces $\ms{W}^*_{t_f}$ way below $\tilde V(0)$ (gray horizontal line), indicating the increased importance of optimal control.

Figure~\figref{protocols}(d) shows $\ms{W}^*_{t_f}$ (symbols) as a function of $u_0$ for fixed $t_f = 0.45\,t_\text{c}$ (red) and $t_f = 1.65\,t_\text{c}$ (blue) together with the theory [\Eqnref{Cmin}, solid lines]. For $t_f < t_\text{c}$, $\ms{W}^*_{t_f}$ is smooth and closely follows the penalty $\tilde V$, both in shape and magnitude. For $t_f > t_\text{c}$, $\ms{W}^*_{t_f}$ is substantially reduced relative to $\tilde V$. In addition, $\ms{W}^*_{t_f}$ develops a kink at $u_0 = 0$, associated with a discontinuous change of the optimal protocol $\lambda^*(t)$ when $\lambda_0 = u_0$ changes sign [see \Figref{protocols}(b)].

The inset in \Figref{protocols}(d) shows the magnitude of the optimal final particle position $|u^*_f|$ for $u_0 = 0$, which serves as the order parameter of the transition (see above). Analogous to the magnetization in an equilibrium spin model as a function of $\beta = 1/(\kb T)$, $|u^*_f|$ vanishes for small $t_f$, where accepting the mean penalty $\tilde V(0)$ is optimal, and becomes finite for $t_f > t_\text{c}$. 
\begin{figure}
	\includegraphics[width = \linewidth]{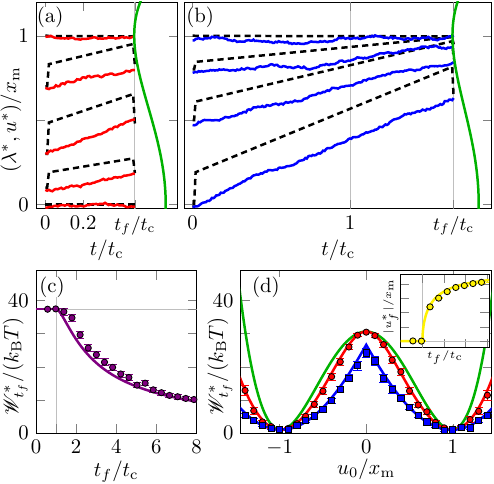}
	\caption{Optimal control protocols for different average initial positions $u_0$ (dashed) and experimental trajectories [solid red in (a), blue in (b)], with process durations: (a) $t_f=0.47\,t_\text{c}$, and (b) $t_f=1.65\,t_\text{c}$. (c) $\ms{W}_{t_f}^*$ at $u_0=0$ as a function of $t_f$ from experiments (markers) and theory (line) for $V_0 = \kappa x_\text{m}^2 = 151\,\kb T$ (d) $\ms{W}_{t_f}^*$ as a function of $u_0$ at $t_f = 0.45\,t_\text{c}$ (red) and $t_f = 1.65\,t_\text{c}$ (blue). Results from experiments (markers) and theory (lines) for $V_0 = \kappa x_\text{m}^2 = 124\,\kb T$. $\tilde V(u_f)$ is shown in green. Inset: $|u^*_f|$ as a function of the control time $t_f$ (yellow) from experiments (markers) and theory (solid line).}\figlab{protocols}
\end{figure}
\paragraph*{FTDPTs in non-equilibrium relaxation.}
The optimal cost $\ms{W}^*_{t_f}$ derived above results from a minimization over trajectories that balance viscous dissipation against a position-dependent energetic penalty. This structure is reminiscent of action functionals arising in large deviation descriptions of stochastic dynamics, where probabilities of trajectories are exponentially weighted according to a probabilistic cost. In particular, the quadratic contribution associated with dissipation and the final penalty term play roles analogous to those governing rare fluctuations in diffusive systems. Notably, this correspondence involves an exchange of boundary conditions: the penalty acting on the final position in the control problem appears as a probabilistic weight of the initial condition in the corresponding relaxation dynamics.

Motivated by this analogy, we map the optimal control problem onto a non-equilibrium relaxation process in which $\ms{W}^*_{t_f}$ acts as a rate function~\cite{Tou09} for rare trajectories. Under this correspondence, the finite time-transition between optimal strategies indeed emerges as a dynamical phase transition in the dominant relaxation pathways. Specifically, we consider a colloidal particle initially prepared at equilibrium in the potential $V(x)$ and subsequently released to diffuse freely. In the regime of weak thermal noise, $\kb T \ll V_0$, and short times $t_f \ll \tau_\text{D} = \gamma x_\text{m}^2/(\kb T)$ compared to the characteristic diffusion time $\tau_\text{D}$ over the distance $x_\text{m}$, the relaxation dynamics is governed by rare events, known as optimal fluctuations.

Within a large deviation framework, detailed in \Secref{ftdpts} of \cite{sup}, these optimal fluctuations are obtained by minimizing the Onsager--Machlup action over stochastic trajectories with fixed boundary conditions $x_0 = x(0)$ and $x_f = x(t_f)$. This procedure closely mirrors the minimization of $\ms{W}_{t_f}$ in the control problem and yields the most probable relaxation pathways connecting $x_0$ and $x_f$.

Their probability scales as $\propto \exp[-\ms{V}_{t_f}(x_f,x_0)/(\kb T)]$, where
\algn{
    \ms{V}_{t_f}(x_f,x_0) &= \frac{\gamma (x_f - x_0)^2}{4 t_f} + V(x_0)\,. \eqnlab{V0t}
}
Maximizing this probability over the initial condition $x_0$ yields the marginal probability $P(x_f,t_f) \propto \exp[-\ms{V}^*_{t_f}(x_f)/(\kb T)]$, characterized by the rate function
\algn{
    \ms{V}^*_{t_f}(x_f) = \min_{x_0} \ms{V}_{t_f}(x_f,x_0)\,.
}
Comparing with \Eqsref{cc3} and \eqnref{Cmin}, the control problem maps directly onto the relaxation problem upon identifying $\ms{W} \leftrightarrow \ms{V}$, $\tilde V \leftrightarrow V$, and $u \leftrightarrow x$, up to a trivial time rescaling $t \to 4(t_f - t)$.
Accordingly, the transition of the optimal control protocol corresponds to a finite-time transition of the dominant relaxation pathway, signaled by a kink in $\ms{V}^*_{t_f}(x)$ at $t^R_\text{c} \approx t_\text{c}/4$, analogous to the kink in $\ms{W}^*_{t_f}(u_0)$ shown in \Figref{protocols}. Such transitions, known as finite-time dynamical phase transitions~\cite{Mei22a,Mei23b}, have been observed primarily in spin systems~\cite{Ent02,Kul07,Erm10,Blo22,Vad24}. By this correspondence, measuring $\ms{W}^*_{t_f}$ in \Figref{protocols}(c) provides not only a direct observation of the control transition but also an indirect signature of an FTDPT.

\begin{figure}
	\includegraphics[width=\linewidth]{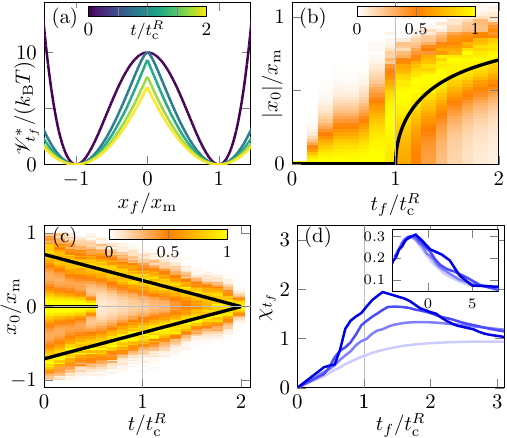}
	\caption{Results of the free-relaxation experiment, post-processed with the reweighting method, 
    for $V_0 = 40\,\kb T$ from $2\cdot10^3$ trajectories. (a) Rate function $\ms{V}^*_{t_f}$ as a function of $x_f$, for different $t_f$. (b) Order parameter density $P(|x_0|,t_f)$ as a function of $t_f$ from experiments (heat map) and $|x^*_0|$ from theory (solid line). $P(|x_0|,t_f)$ is scaled to unity at each $t_f$ to improve visibility. (c) Trajectory density of relaxation trajectories that reach $x_f=0$ at different times $t_f$ from experiment (heat map), combined for $t_f=0.55t^R_\text{c}$ and $t=2t^R_\text{c}$. Solid lines show the theoretical predictions. (d) Susceptibility of the order parameter $x_0$ as a function $t_f$ at $V_0=10\,\kb T$, $20\,\kb T$, $30\,\kb T$, and $40\,\kb T$ (from light to dark). Inset: Scaling behavior of $\chi_t$.}\figlab{ratefunc_relax}
\end{figure}
To probe this relaxation transition directly, we now turn to an experiment that accesses $\ms{V}^*_{t_f}$. The main practical difference between $\ms{W}^*_{t_f}$ and $\ms{V}^*_{t_f}$ is that the latter characterizes the probability of rare relaxation events, while the former corresponds to a statistical average. Observing such rare events requires $\propto \exp[\ms{V}^*_{t_f}(x_f)/(\kb T)] \gg 1$ measurements, which poses a serious experimental challenge. To overcome this, we employed an importance-sampling technique commonly used in numerical simulations~\cite{Buc04}: we recorded approximately $2\cdot10^3$ trajectories of duration $0.5\,\unit{\second}\approx \tau_\text{D}/4$ of a freely diffusing Brownian particle with an arbitrary initial distribution (see \Secref{Aexpmeth} in \cite{sup}), and subsequently reweighted them such that the initial distribution corresponds to the Boltzmann weight $\propto \exp[-V(x)/(\kb T)]$. This way we circumvent the need to include an equilibration period in the experiment. Figure~\figref{ratefunc_relax}(a) shows $\ms{V}^*_{t_f}(x_f) \sim -\kb T \log P(x_f,t_f)$ obtained from reweighted experimental trajectories at times below and above $t^R_\text{c}$. The emergence of a kink at $x=0$ for $t>t^R_\text{c}$ is clearly visible.

We define an order parameter for the FTDPT by the absolute value $|x_0^*|$ of the minimizer $x_0^* = \text{argmin}_{x_0} \ms{V}_{t_f}(x=0,x_0)$. For $\kb T \ll V_0$, the probability density $P(|x_0|,t_f)$ of initial positions leading to $x_f=0$ at time $t_f$ is sharply peaked around $|x_0^*|$. Figure~\figref{ratefunc_relax}(b) shows $P(|x_0|,t_f)$ as a heat map, providing clear evidence of the transition at $t=t^R_\text{c}$.

The FTDPT in the relaxation experiment originates from a sudden change in the dominant relaxation pathway. Figure~\figref{ratefunc_relax}(c) shows the density of trajectories conditioned on reaching $x_f=0$ at time $t_f$. For $t_f<t^R_\text{c}$, the density is concentrated around an optimal fluctuation that starts at $x_0=0$ and effectively remains there. Although rare, this is exponentially more likely than trajectories starting near $\pm x_\text{m}$ and reaching $x_f=0$ within the short time window. For $t_f>t^R_\text{c}$, by contrast, trajectories originate near $\pm x_\text{m}$ and relax towards $x_f=0$ along approximately linear paths (black), in analogy with the optimal control solution $u^*(t)$.

Finally, to characterize fluctuations of $|x_0^*|$ near the critical point, we analyzed the susceptibility $\chi_{t}$ extracted from experimental data [Fig.~\figref{ratefunc_relax}(d)]. As $V_0/\kb T$ increases, $\chi_{t}$ develops a growing peak that shifts towards $t^R_\text{c}$ (gray line). When suitably rescaling $\chi_{t_f}$ and $\delta t = (t_f-t^R_\text{c})/t_f$ (inset) with $(V_0/\kb T)^{1/2}$, the curves collapse onto a scaling function, suggesting that the transition are characterized by mean-field critical exponents. By contrast, the fluctuations in the optimal control experiment show no such critical behavior.

\paragraph*{Conclusion.}
We theoretically and experimentally showed that optimal control of particles in structured environments exhibits finite-time transitions. The transition arises from a trade-off between dissipation along longer trajectories that avoid unfavorable regions and the energetic penalty of crossing them. This competition yielded distinct optimal protocols depending on control time. We further showed that these transitions are directly linked to finite-time dynamical phase transitions in non-equilibrium relaxation, reflecting a switch between dominant pathways at short and long times. In the companion paper~\cite{Mei26c}, we demonstrate that these effects extend to a broader class of control problems. Our findings suggest that abrupt transitions are generic in mesoscopic control in structured environments and that optimal-control experiments provide access to dynamical phase transitions without sampling exponentially rare trajectories.
\paragraph*{Acknowledgments.} JM thanks John Bechhoefer for pointing out the relevance of Refs.~\cite{Kap05a,Kap05b}. CB acknowledges funding by the Deutsche Forschungsgemeinschaft, SFB 1432 (425217212) and the  ERC AdG.Grant No.101141477.

\paragraph*{Data availability.}
The data that support the findings of this manuscript will be deposited in a Zenodo repository upon publication. In the meantime, they are available from the authors upon request.

\end{document}

\title{Supplemental Material to: Finite-time transitions in optimal control and non-equilibrium relaxation}

\author{Jan Meibohm}
\affiliation{Institute for Physics and Astronomy, Technische Universit\"at Berlin, Hardenbergstra\ss{}e 36, 10623 Berlin}
\author{Samuel Monter}
\affiliation{Faculty of Physics, University of Konstanz, Konstanz, Germany}
\author{Sarah A. M. Loos}
\affiliation{Max Planck Institute for Dynamics and Self-Organization, G\"ottingen}
\author{Clemens Bechinger}
\affiliation{Faculty of Physics, University of Konstanz, Konstanz, Germany}

\begin{abstract}
In this Supplemental Material, we provide details about the derivations in the main text and describe the experiments. We first derive the noise-averaged potential $\tilde V$ and the optimal trajectories for the protocol and the mean position. We then show how to compute the critical time and briefly describe the connection between rate function in non-equilibrium relaxation and the cost function in optimal control. Finally, we provide details about the experimental setup and the measurement procedures, including the importance-sampling method that we used to measure the rate function in the experiment.
\end{abstract}

\maketitle
%
% Supplemental Material numbering: S1, S2, ... so that SM equations,
% figures and sections cannot be confused with those of the main text.
\setcounter{equation}{0}\renewcommand{\theequation}{S\arabic{equation}}
\setcounter{figure}{0}\renewcommand{\thefigure}{S\arabic{figure}}
\setcounter{table}{0}\renewcommand{\thetable}{S\arabic{table}}
\setcounter{section}{0}\renewcommand{\thesection}{\Roman{section}}
\section{Noise-averaged potential}\seclab{potential}
In this section, we derive the noise-averaged potential $\tilde V(u_f)$ from the bare potential $V(x)$. Due to the linear Langevin dynamics [\MEqnref{Langevinequ}], we know that $x_f$ is Gaussian with stationary variance $\langle x_f^2\rangle-\langle x_f\rangle^2 = \kb T/\kappa$. For the double-well form [\MEqnref{Vx}]
\algn{
	V(x) = \frac{V_0}{4}\left(\frac{x^2}{x_\text{m}^2} - 1\right)^{2}\,,
}
we thus have
\algn{\eqnlab{potential}
	\langle V(x_f)\rangle = \frac{V_0}{4}\left(\frac{\langle x_f^4\rangle }{x_\text{m}^4} - 2\frac{\langle x_f^2\rangle }{x_\text{m}^2} + 1\right)\,.
}
To obtain $\tilde V(u_f)$, we need to express $\langle x_f^2\rangle$ and $\langle x_f^4\rangle$ in terms of $\langle x_f\rangle = u_f$. Using the Gaussian properties of $x_f$, we find
\algn{
	\langle x_f^2\rangle 	=& u_f^2 + \frac{\kb T}{\kappa}\,,\\
	\langle x_f^4\rangle 	=& u_f^4  + 6\left(\frac{\kb T}{\kappa}\right) u^2_f + 3 \left(\frac{\kb T}{\kappa}\right)^2\,.
}
We thus obtain for the noise-averaged potential $\langle V(x_f) \rangle = \tilde V(u_f)$ in \Eqnref{potential}
\begin{equation}
	\tilde V(u_f) = V(u_f) + 
		\frac{V_0}{4}\left(\frac{\kb T}{\kappa x_\text{m}^2}\right)\left(\frac{6 u_f^2}{x_\text{m}^2}- 2 + \frac{3\kb T}{\kappa x_\text{m}^2}\right)\,.
\eqnlab{potential2}
\end{equation}
Equation~\eqnref{potential} shows that $u_f$ experiences the noise-averaged penalty $\tilde V(u_f)$ that is in general different from the penalty of individual trajectories $V(x_f)$. For weak noise, $\kb T\ll V_0,\kappa x_\text{m}^2$, however, the second term in \Eqnref{potential2} can be neglected and we have $\tilde V(u_f)\approx V(u_f)$, as used in the main text.
\section{Derivation of optimal protocols and optimal work}\seclab{optimalprot}
Starting with \MEqnref{cost} we perform an integration by part to remove the time derivative of $\lambda(t)$. We then introduce the Lagrange multiplier $\mu(t)$ to enforce that $u(t)$ must satisfy the noise-averaged dynamics
\algn{
	\dot u(t) =  -\tau_\text{p}^{-1}[u(t)-\lambda(t)]\,.
}
This leads us to
\algn{
	\ms{W}_{t_f}[u(t),\mu(t),\lambda(t)] = \int_0^{t_f}\!\!\ed t \, \left\{\mu \dot u - \tau_\text{p}^{-1}[\mu(t)-\kappa\lambda(t)][\lambda(t)-u(t)]\right\} + \kappa\lambda_f\left(\frac{\lambda_f}2 - u_f\right) - \kappa\lambda_0\left(\frac{\lambda_0}2 - u_0\right) + \tilde V(u_f)\,.
}
To simplify the calculations that follow, we add
\algn{
	0 = \frac{\kappa}2(u_f^2-u_0^2) - \frac{\kappa}{\tau_\text{p}}\int_0^{t_f}\!\!\! \ed t\ u(t)[\lambda(t)-u(t)]\,,
}
which gives 
\algn{
	\ms{W}_{t_f}[u(t),\mu(t),\lambda(t)] = \int_0^{t_f}\!\!\ed t \, \left\{\mu \dot u - \tau_\text{p}^{-1}[\kappa u(t)-\kappa\lambda(t)+\mu(t)][\lambda(t)-u(t)]\right\} + \frac{\kappa}{2}\left(\lambda_f - u_f\right)^2 - \frac{\kappa}2\left(\lambda_0 - u_0\right)^2 + \tilde V(u_f)\,.
}
Peforming a variation over $\lambda(t)$ and setting $\delta_{\lambda(t)}\ms{W}_{t_f}[u(t),\lambda(t),\mu(t)]\big|_{\lambda(t) = \lambda^*(t)} =0$, we find
\algn{\eqnlab{Alamsol}
	\lambda^*(t) = \frac1{2\kappa }\left[\mu(t) + 2\kappa u(t)\right]\,,\quad \lambda_{f,0} = u_{f,0}\,.
}
We thus have for $\ms{W}_{t_f}[u(t),\mu(t)] = \ms{W}_{t_f}[u(t),\mu(t),\lambda^*(t)]$,
\algn{\eqnlab{newcost}
	\ms{W}_{t_f}[u(t),\mu(t)] = \int_0^{t_f}\!\!\ed t \, \left\{\dot u(t)\mu(t) - \ms{H}[u(t),\mu(t)]\right\} + \tilde V(u_f)\,,
}
with the simple quadratic control Hamiltonian
\algn{\eqnlab{Hamquad}
	\ms{H}[u(t),\mu(t)] = \frac{\mu^2(t)}{4\gamma}\,.
}
Performing the variation over $u(t)$ and $\mu(t)$ while holding $u_f$ and $u_0$ fixed and requiring stationarity $\delta \ms{W}=0$, we obtain Hamilton's equations of motion
\algn{\eqnlab{hameom}
	\dot u^*(t) = \frac{\partial \ms{H}}{\partial \mu(t)}=\frac{\mu^*(t)}{2\gamma}\,,	\qquad \dot \mu^*(t) = 0\,,
}
from which we obtain the solutions
\algn{
	\mu^*(t) = \mu^*(0)\,,\qquad u^*(t) = \frac{\mu^*(0)t}{2\gamma} + u_0\,.
}
Evaluating at $t=t_f$ and written in terms of $\mu^*(t_f)=\mu^*(0)$, we find
\algn{\eqnlab{mueqs}
	\mu^*(t_f)=\mu^*(0) = 2\kappa(u_f - u_0)\frac{\tau_\text{p}}{t_f}\,,
}
which gives the optimal solutions
\algn{
	u^*(t) =& (u_f - u_0)\frac{t}{t_f} + u_0\,,\\
	\lambda^*(t) =& t_f^{-1}(u_f - u_0)\left(t+\tau_\text{p}\right) + u_0\,,
}
We observe that $u^*(t)$ is continuous with $u^*(0) = u_0$ and $u^*(t_f)=u_f$. For the optimal protocol $\lambda^*(t)$, by contrast, we have the limits $\lambda^*(0^+) = u_0+(u_f - u_0)\tau_\text{p}/t_f$ and $\lambda^*(t_f^-) = u_f+(u_f - u_0)\tau_\text{p}/t_f$. This shows that $\lambda^*(t)$ has jumps of equal, size $\lambda^*(0^+)-u_0 = \lambda^*(t_f^-)-u_f = (u_f - u_0)\tau_\text{p}/t_f$, at the start and end points as depicted in \MFigref{protocols}{}.

To derive the optimal work $\ms{W}^*_{t_f}$ [\MEqnref{cc3}], we rewrite
\algn{\eqnlab{relation}
	\int_0^{t_f}\!\!\ed t \, \dot u(t)\mu(t) = \frac12\int_0^{t_f}\!\!\ed t \, \left[\dot u(t)\mu(t) - u(t) \dot \mu(t)\right]+ \frac12\left[u_f\mu(t_f) - u_0\mu(0)\right]
}
Using the Hamiltonian equations~\eqnref{hameom}, it is now simple to show that
\algn{
	\frac12\int_0^{t_f}\!\!\ed t \, \left[\dot u^*(t)\mu^*(t) - u^*(t) \dot \mu^*(t)\right] = \int_0^{t_f}\!\!\ed t \, \ms{H}[u^*(t),\mu^*(t)]\,.
}
Substituting these relations into \Eqnref{newcost} gives for $\ms{W}_{t_f}(u_0,u_f) = \ms{W}_{t_f}[u^*(t),\mu^*(t)]$:
\algn{
	\ms{W}_{t_f}(u_0,u_f) =& \frac12\left[u_f\mu^*(t_f) - u_0\mu^*(0)\right] +\tilde V(u_f)\,,\nn\\
	 =& \frac{\gamma(u_f-u_0)^2}{t_f} +\tilde V(u_f)\,, \eqnlab{Wtf}
}
where we used \Eqsref{mueqs} in the final step. This completes the derivation of \MEqnref{cc3}.
\section{Derivation of the critical time}\seclab{tcrit}
In order to establish from \Eqnref{Wtf} [\MEqnref{cc3}] the existence of a finite-time transition of the control protocol, we need to determine the stability of the zero-control solution $u_0=u_f=0$. To this end, we evaluate $\ms{W}^*_{t_f}$ at $u_0=0$ and expand to second order in $u_f$. Assuming that $\tilde V(u_f)$ is even, $\tilde V(u_f) = \tilde V(-u_f)$, this gives
\algn{\eqnlab{Wexp}
	\ms{W}_{t_f}(0,u_f) &\sim \tilde V(0) + \left[\frac{\gamma}{t_f} +\frac{\tilde V''(0)}2\right]u_f^2\,.
}
For short times, $t\ll \tau_{\text{p}}$ the factor in front of the $\propto u_f^2$ term is positive. This implies that starting from $u_0=0$, the optimal solution will end at $u_f=0$, the minimum of the cost function. Hence, for short enough times, the zero-control protocol $\lambda^*(t)=u^*(t)=0$ is optimal. However, the $\propto u_f^2$ factor in \Eqnref{Wexp} changes sign when $t_f$ exceeds the critical time $t_\text{c}$, given by
\algn{
	t_\text{c} &= \frac{2\gamma}{-\tilde V''(0)}\,,
}
which is \MEqnref{tcrit}.
\section{Rate functions and FTDPTs}\seclab{ftdpts}
In the weak noise-limit, the large deviations of the relaxation process that follows the quench are dominated by so-called optimal fluctuations, describing the most likely ways in which the system relaxes. These optimal fluctuations minimize the Onsager-Machlup action
\algn{\eqnlab{action}
	\ms{V}_{t_f}[x(t),\dot x(t)] = \int_0^{t_f}\!\!\ed t\ \ms{L}[x(t),\dot x(t)] + V(x_0)\,,
}
with the Lagrangian $\ms{L}[x(t),\dot x(t)] = \gamma \dot x^2(t)/4$. We turn to a Hamiltonian description by performing the Legendre transform
\algn{
	\ms{H}_\text{R}[x(t),p(t)] =& p(t)\dot x(t)- \ms{L}[x(t),\dot x(t)]\,, \\
	p(t) =& \frac{\partial \ms{L}}{\partial \dot x}\,,
}
yielding the Hamiltonian
\algn{
	\ms{H}_\text{R}[x(t),p(t)] = \frac{p^2(t)}{\gamma}\,.
}
Rewriting \Eqnref{action} in terms of $\ms{H}_\text{R}$ then yields
\algn{\eqnlab{Vtf}
	\ms{V}_{t_f}[x(t),p(t)] = \int_0^{t_f}\!\!\ed t \left\{p(t)\dot x(t) - \ms{H}_\text{R}[x(t),p(t)]\right\} + V(x_0)\,,
}
with the particle position trajectory $x(t)$, $x(t_f)=x_f$ and the ``conjugate momentum'' $p(t)$~\cite{Mei22a}. The latter represents the effect of the thermal noise in the weak-noise limit.

Comparison with \Eqsref{newcost} and \eqnref{Hamquad}, shows that these formulations are identical up to a factor of $4$. Using the same steps to minimize $\ms{V}_{t_f}[x(t),p(t)]$, we arrive at the rate function for fixed $x_0$ and $x_f$:
\algn{
	\ms{V}_{t_f}(x_0,x_f) &= \frac{\gamma(x_f-x_0)^2}{4t_f} + V(x_0)\,, \eqnlab{AV0t}
}
in analogy with the stochastic work $\ms{W}_{t_f}$ in \Eqnref{Wtf} [\MEqnref{cc3}], where the initial distribution $V$ plays the role of the average final penalty $\tilde V(u_f)$. In particular, after the critical time $t^R_\text{c}\approx t_\text{c}/4$, the rate function $\ms{V}^*_t(x_f) = \min_{x_0}\ms{V}_t(x_f,x_0)$, obtained after minimizing over the initial position $x_0$, attains a kink analogous to the kink in $\ms{W}^*_{t_f}(u_0)$.
\section{Details of experiments}\seclab{Aexpmeth}
\subsection{Optical tweezers setup}\seclab{Aexpsetup}
A 532 nm laser (Coherent Verdi V2) is used as the coherent light source for the optical tweezers setup. 
The beam intensity can be modulated and its angle deflected using an acousto-optic deflector (AOD, AA Opto-Electronic DTSXY-400). 
A telescope composed of 2 inch lenses directs the deflected beam to the back aperture of the microscope objective (Olympus Apochromat MPLAPON-Oil 100x, NA = 1.45), accommodating deflection angles of up to \textpm~40~\unit{\milli\radian}. 
The objective serves both for optical trapping and imaging. 
The laser power is adjusted such that, at the highest AOD setting used in experiments, it reaches a maximum of 160~\unit{\milli\watt} immediately before the back aperture.
Microscopy videos are recorded with a digital camera (Basler ace 2 a2A3840-45umPRO) at a frame rate of 250~\unit{\hertz} in the optimal control and 400~\unit{\hertz} in both relaxation experiments, with the region of interest adapted to the specific experimental requirements. 
The sample temperature is controlled by resistive heating (Okolab H401-T-Penny) applied to both the sample stage and the objective, and is maintained at 25 \textpm 0.05~\unit{\celsius} throughout all experiments.
This configuration allows the laser focus to be shifted within the imaging plane by up to \textpm 19~\unit{\micro\meter}. 
To suppress AOD-induced artifacts, a zero-intensity interval is inserted between consecutive positions in the control sequence, resulting in a pattern such as (position 1, zero, position 2, zero, ...).\\
\indent Calibration is required for two reasons, the first being that it enables the conversion between spatial positions within the camera's field of view and the corresponding control frequencies applied to the AOD. 
Second, the AOD exhibits a frequency-dependent diffraction efficiency, which induces systematic variations in the trap stiffness.
We use the same calibration method as in another recent experimental study \cite{Mon26}.
Calibration establishes the mapping between AOD control frequencies and trap positions by sampling a grid across the field of view and assigning time-averaged particle positions to the corresponding laser focus locations. A continuous mapping is then obtained via interpolation.
To account for spatial variations in trapping strength, the trap stiffness $\kappa_i$ is characterized across the field of view and normalized to its mean value in each spatial direction $\langle\kappa_i\rangle$. An correction factor is defined for each frequency pair as
\begin{equation}
    c(f_x,f_y) = \frac{\kappa_x/\langle \kappa_x\rangle + \kappa_y/\langle \kappa_y\rangle}{2}.
\end{equation}
A smooth correction function $C(f_x,f_y)$ is obtained by fitting a third order polynomial to $c(f_x,f_y)$. $C(f_x,f_y)$ is then used to rescale the applied AOD amplitude according to
\begin{equation}
    A_\mathrm{apply} = \frac{A_\mathrm{set}}{0.5 + 0.5\cdot C(f_x,f_y)}.
\end{equation}
Using this approach, spatial variations in trap stiffness are reduced from \textpm 20~\unit{\percent} to \textpm 5 ~\unit{\percent}.\\
Sample suspensions are prepared by dispersing spherical silica micro particles (diameter $2.73 \pm 0.12~\unit{\micro\meter}$, microParticles GmbH) in a water-glycerol mixture at a volumetric ratio of 1:1. 
Measurement cells are assembled by filling glass capillaries (inner diameter 100~\unit{\micro\meter}, width 1~\unit{\milli\meter}; CM Scientific) with the prepared suspensions. The open ends of the capillaries are sealed using a combination of wax and epoxy resin to prevent evaporation. 
The samples are allowed to equilibrate within the measurement setup for a minimum of 60 minutes prior to data acquisition.

\subsection{Measurement Procedures}\seclab{Aexpmeas}
\subsubsection{Optimal Control Experiment}\seclab{Aexpcontrol}
After trapping a particle near the midplane of the sample cell (to avoid wall-interactions) a calibration measurement is performed as described in ~\secref{Aexpsetup}.
To test the optimal control protocols for varying initial positions $u_0$ and protocol durations $t_f$, the trap center $\lambda(t)$ is moved according to the corresponding optimal protocol given in \MEqnref{lamsol}. Each experiment consists of both forward and backward implementations of this protocol. This avoids a continuous drift of the trap across the sample, which would be incompatible with the finite deflection range of the AOD and the limited field of view of the camera. In addition, averaging over forward and backward realizations reduces variations in trap stiffness are not fully accounted for by the calibration procedure.
Before, between and after each forward and backward execution, the trap is held fixed for 3~\unit{second} to ensure that each realization starts from equilibrium. To obtain statistically significant averages, each forward/backward experiment is repeated at least 40 times, yielding 80 trajectories per protocol.\\
\indent From the measured trajectories, we evaluate the stochastic work following \cite{Cro98} as
\begin{equation}
    W = \sum_{i=1}^{N} U[x(t_i),\lambda(t_{i+1})] - U[x(t_i),\lambda(t_i)]\label{eq:Wcrooks}
\end{equation}
where time is discretized into N intervals of duration $t_f/(N-1)=0.004~\unit{\second}$ defined by the acquisition rate of the camera.

\subsubsection{Free relaxation experiment and reweighting}\seclab{Aexpfreerelax}
For the relaxation experiments, we first localized a single particle in the optical trap. We then repeatedly turned the trap on and off for time periods $t_\text{on} = t_\text{off} = 0.5\,\unit{\second}$, respectively. During the ``off'' time, we observed free diffusion of the particle with diffusion constant $D=\kb T/\gamma\approx 0.022\,\unit{\micro\meter^2}/\unit{\second}$ and friction coefficient $\gamma\approx 0.18\,\unit{\mu Ns/m}$. After the time period $t_{\mathrm{off}}$, the trap was turned back on to prevent diffusion of the particle out of the field of view and to prevent the particle from settling outside the focal plane.

To compute the rate function $\ms{V}_{t_f}^*(x_f)$ shown in \MFigref{ratefunc_relax}{}, we considered the probability density
\algn{\eqnlab{probintegral}
	P(x_f,t_f) = \int_{-\infty}^\infty\!\!\!\ed x_0\ P(x_f,t_f|x_0) P(x_0)\,,
}
where $P(x_0)$ denotes the initial probability density at $t=0$ and $P(x_f,t_f|x_0)$ the transition probability from $x_0$ at time $t=0$ to $x_f$ at time $t_f$. The initial distribution before the quench is given by the Boltzmann distribution
\algn{\eqnlab{boltzmann}
	P(x_0) = \frac{\ee^{-\frac{V(x)}{\kb T}}}{Z}\,,\qquad Z = \int_{-\infty}^\infty\!\!\!\ed x\ \ee^{-\frac{V(x)}{\kb T}}\,.
}
In the weak-noise limit $\kb T\ll V_0$, the rate function $\ms{V}^*_{t_f}$ is obtained from the large-deviation form
\algn{
	P(x_f,t_f) \propto \ee^{-\frac{\ms{V}^*_{t_f}(x_f)}{\kb T}}\,.
}
A similar form applies for the transition probability
\algn{
	P(x_f,t_f|x_0) \propto \ee^{-\frac{\ms{V}_{t_f}(x_f|x_0)}{\kb T}}\,.
}
From a saddle-point approximation of \Eqnref{probintegral} we now find
\algn{\eqnlab{minrel}
	\ms{V}_{t_f}^*(x_f) = \min_{x_0}\left\{\ms{V}_{t_f}(x_f|x_0) + V(x_0)\right\}\,.
}
With normal sampling, it is unfortunately challenging to obtain $\ms{V}^*_{t_f}(x_f)$ with satisfactory accuracy. Instead, we obtained $\ms{V}^*_{t_f}(x_f)$ by measuring the transition probability $P(x_f,t_f|x_0)$ experimentally and using \Eqnref{minrel}. 

To this end, we note that translational and inversion invariance of free diffusion implies $P(x_f,t_f|x_0) = P(|x_f-x_0|,t_f|0)$, so $\ms{V}_{t_f}(x_f|x_0) = \ms{V}_{t_f}(|x_f-x_0||0)$. Hence,
\algn{\eqnlab{minrel2}
	\ms{V}_{t_f}^*(x_f) = \min_{x_0}\left\{\ms{V}_{t_f}(x_0|0) + V(x_0+x_f)\right\}\,.
}
This implies that to obtain $\ms{V}_{t_f}^*(x_f)$, we need to measure $\ms{V}_{t_f}(x_0|0)$ as a function of $x_0$ and reweight the result with $V(x)$, as described by \Eqnref{minrel2}.

To do this efficiently, we again use translational invariance to shift all measured trajectories to $x_0=0$, so that $P_s(x_0)=\delta(x_0)$ in \Eqnref{probintegral}. For the probability density $P_s(x_f,t_f)$ of the shifted trajectories, we then find
\algn{
	P_s(x_f,t_f) = P(x_f,t_f|0)\,.
}
We can therefore compute $\ms{V}_{t_f}(x_0|0)$ in \Eqnref{minrel2} by
\algn{
	\ms{V}_{t_f}(x_0|0) \sim -\kb T \log P_s(x_f,t_f)|_{x_f = x_0}\,.
}
To further increase the sample size, we used the Markov property of the diffusion process to split long trajectories into snippets of the required time window $t_f$. This enabled us to obtain trajectory samples of sufficient size ($\approx 10^6$) from about $2\cdot10^3$ Brownian trajectories of $t_\text{off}= 0.5$\,s length to obtain an accurate estimate of $\ms{V}_{t_f}(x_0|0)$. We then used \Eqnref{minrel2} to determine the rate function $\ms{V}_{t_f}^*(x_f)$ as shown in \MFigref{ratefunc_relax}{(a)}.

Similarly, to determine the order-parameter density $P(|x_0|,t_f) = P(|x_0|,0|0,t_f)$ shown in \MFigref{ratefunc_relax}{(b)}, we need to compute
\algn{
	P(|x_0|,0|0,t_f) 	=& \frac{P(0,t_f|x_0)P(|x_0|)}{P(0,t_f)}\nn\,,\\
				\propto& \exp\left[-\frac{\ms{V}_{t_f}(0|x_0) + V(x_0)-\ms{V}^*_{t_f}(0)}{\kb T}\right]\,,
}
where we used $V(x_0) = V(-x_0)$. Using again the invariance properties of Brownian motion, we have $\ms{V}_{t_f}(0|x_0) = \ms{V}_{t_f}(x_0|0)\sim  -\kb T \log P_s(x_f,t_f)|_{x_f = x_0}$. This way, we calculated $P(|x_0|,0|0,t_f)$ in \MFigref{ratefunc_relax}{(b)} from the experimental measurements.

Finally, to obtain the density of trajectories that end in $x_f=0$ at $t_f$, shown in \MFigref{ratefunc_relax}{(c)}, we note that $P(0,t_f|x_0) = P(x_0,t_f|0) = P_s(x_f,t_f)|_{x_f = x_0}$. This implies that
\algn{
	P(0,t_f) = \int_{-\infty}^\infty\!\!\!\ed x_0\ P_s(x_0,t_f)P(x_0)\,,
}
for the initial equilibrium distribution $P(x_0)$ in \Eqnref{boltzmann}. We now note that for a uniform initial density $P_u(x_0) = 1/\ell$ over an arbitrary interval $x_0\in[-\ell/2,\ell/2]$, we have
\algn{
	P(0,t_f|x_0)P_u(x_0) = P(0,t|x_0)\ell^{-1}\,,
}
which is proportional to $P_s(x_f,t_f)|_{x_f = x_0}$. This means that $P_s(x_f,t_f)|_{x_f = x_0}\ell^{-1}$ can be interpreted as $P(0,t_f|x_0)P_u(x_0)$, which is the joint probability of sampling initial positions $x_0$ from a uniform distribution and ending in $x_f=0$ at $t_f$.  Hence, to obtain the trajectory density corresponding to $P(x_0)$ in \Eqnref{boltzmann}, we need to reweight the experimental trajectories contributing to $P_s(x_f,t_f)|_{x_f = x_0}$ according to their initial value $x_0$ by $M = P(x_0)/P(0)$ to transform the uniform distribution $P_u(x_0)$ into $P(x_0)$. For the resulting, reweighted $P^*_s(x_0,t_f)$, we then have $P^*_s(x_0,t_f) = P(0,t_f|x_0)P(x_0)$. The trajectory density of this reweighted ensemble is shown in \MFigref{ratefunc_relax}{(c)} for two different $t_f$.